\documentclass[12pt]{article}
\usepackage[english]{babel}
\usepackage{epsf}
\usepackage{pazh}
\usepackage{amsmath}	% Advanced maths commands
\usepackage{amssymb}	% Extra maths symbols
\newcommand{\kms}{\,km\,s$^{-1}$}

\begin{document}
\title{\bf Luminosity source in supernova ASASSN-15nx with long linear light curve}
\author{ 
N. N. Chugai
}
\affil{
Institute of Astronomy of Russian Academy of Sciences, Moscow\\ 
}

\vspace{2mm}
%\received{\today}

\sloppypar 
\vspace{2mm}
\noindent


\noindent
%{\bf PACS codes:\/} ?????

\vfill
\noindent\rule{8cm}{1pt}\\
{$^*$ email: $<$nchugai@inasan.ru$>$}

\clearpage

\begin{abstract}
\centerline{\bf Abstract}
\vspace{0.8cm}

The available spectra of the anomalous supernova ASASSN-15nx permit us to 
rule out the radioactivity and circumstellar interaction as the luminosity source.
I propose an alternative mechanism for the ASASSN-15nx luminosity based on the 
interaction of the neutron star rotating magnetosphere with the gravitationally 
bound material of the envelope ejected by the shock wave. In the regime of 
stationary accretion the rotational frequency decreases exponentially with time, 
which could account for the linearity of the light curve. The modelling of the 
light curve at the stage of the luminosity rise in combination with the expansion 
velocity implies the low mass of ejecta, $\sim$1\msun. The profile 
of the [O\,I] 6300, 6364\,\AA\ doublet indicates the asphericity of the oxygen 
distribution, which in turn suggests the aspherical explosion.

\end{abstract}

\section{Introduction}

The anomalous type II supernova ASASSN-15nx (Bose et al. 2018) with the maximal 
magnitude $M_V \approx -20$ demonstrates unusual light curve that is characterized by 
a perfect, as authors emphasise, linear decline (2.5 mag per 100 d) during long period 
of 250 days. This peculiarity has no counterpart among SNe~II, including SNe~IIL.
Such a light curve cannot be an outcome of the diffusion cooling of the explosion 
energy: the long-lasting source of energy is required. In the original paper authors propose 
two possibilities: (i) 1.6\msun\ of the radioactive $^{56}$Ni in the 2\msun\ ejecta; 
(ii) the energy released in the circumstellar (CS) interaction (Bose et al. 2018).
Generally, both 
mechanisms can operate in SNe~II and either could be a contender for the power source 
in ASASSN-15nx.

A conjecture on the nature of the energy source in ASASSN-15nx, however, requires verification.
A top priority is the examination of possible spectral effects that accompany 
a certain mechanism. Note that the spectral verification 
was not a subject of the original paper. It is essential therefore to make up 
for the lack of this study, which is one of goals of this paper addressed in the 
next section. As will become clear, the spectra  
reveal serious difficulties both for radioactive mechanism and for the CS interaction.
I will propose an alternative mechanism that, under minimum of assumptions, is 
able to account for the luminosity and the linear light curve of ASASSN-15nx.

Below I use the explosion moment and distance following Bose et al (2018).

\section{Interpretation of ASASSN-15nx spectra}

\subsection{Lines identification}

Spectra of ASASSN-15nx taken between days 53 and 262 after the explosion (Bose et al. 2018) 
look like a set of emission lines on the quasi-continuum composed by numerous metal lines 
(Fig. 1). The dominant emission line are H$\alpha$ (detail $b$), triplets of 
Ca\,II 8600\,\AA\ ($g$) and O\,I 7774\,\AA, ($e$), doublets of Na\,I 5892\,\AA\ ($a$), and  
[Ca\,II] 7300\,\AA\ ($d$). Late time spectra show doublet [O\,I] 6300, 6364\,\AA.
All these lines lie in the red part of the spectrum that will be addressed below.
Some unusual features of the spectra has been already noted by Bose et al. (2018), 
particularly, two-component structure of O\,I 7774 \AA\ profile on day 53. Yet several 
interesting detailes, central to understanding of the energy source, remained beyond the scope.

Since the line identification is hampered by the non-local scattering in line blends as well as 
the Thomson scattering we rely on the synthetic spectrum that takes into account both effects.
The model suggests spherical envelope with the homologous expansion, $v = r/t$. The line intrinsic 
emissivity is set in the parametric form $j = j_0/[1 + (v/v_0)^k]$. The line Sobolev optical depth 
is set by the similar distribution but with a different power index $k$.
At the certain age parameters $v_0$ and $k$ are the same, except for oxygen lines. The continuum emissivity 
is distributed in the same way. The electron distribution is set assuming the recombination origin of H$\alpha$ as $n_e \propto \sqrt{j(\mbox{H}\alpha)}$.

The Thomson optical depth on day 53 is estimated from the H$\alpha$ with the spectrum 
fluxed on the bases of the photometric data (Bose et al. 2018). The inferred H$\alpha$ luminosity at this age 
is $L = 7\times10^{39}$ erg\,s$^{-1}$. When combined with the H$\alpha$ 
profile this value results in the 
Thomson optical depth $\tau_{\sc T} = 1.3$. The found electron distribution is used for the 
synthetic spectrum on day 53. The spectrun is computed by the Monte Carlo technique. Note that we does not 
compute quasi-continuum produced by the emission of numerous metal lines; instead we adopt 
smooth continuum. This proviso should be taken into account when comparing synthetic and observational 
spectra.

I start with the two-component profile of  O\,I 7774\,\AA\ on day 53 that was 
highlighted by Bose et al. (2018). The importance of this issue is emphasised by a possible relation with the oxygen 
asymmetry addressed below in a separate section.
In fact, the  two-component structure of  O\,I 7774\,\AA\ on day 53 has nothing to do with 
the asymmetry;
instead the red component should be attributed to Mg\,II 7877, 7896 \AA\ emission lines. 
The presence of these lines is apparent also on day 87 as well (emission $e$); the 
same spectrum shows Mg\,II 8214, 8235\,\AA\ (emission $f$). Besides, Mg\,II 9218, 7896\,\AA\ together with O\,I 9261, 9266\,\AA\ constitute the emission 9250\,\AA\ (band $h$). All the identified lines of Mg\,II and O\,I with high contrast were identified earlier in SN~Ibn SN~2006jc (Chugai 2009).

The ASASSN-15nx spectra at all phases show He\,I 7065 \AA\ line (Fig. 1). This suggests that the He\,I 5876 \AA\ line contributes to the Na\,I 5892 \AA\ emission with the relative 
intensity $I(5876)/I(7065)$ in the range of 1...~2 judging by spectra of SN~2006jc 
(Anupama et al. 2009). The He\,I 5876 \AA\ is included in all the synthetic spectra shown in Fig. 1.

\subsection{The absence of [Co\,III] 5890\,\AA\ line}

The radiactive mechanism for the ASASSN-15nx luminosity suggests 1.6\msun\ of 
$^{56}$Ni in the low mass envelope, 2\msun\ (Bose et al. 2018). The direct implication of 
this model should be a strong emission line of [Co\,III] 5890\,\AA\ at 70-150 days 
likewise in spectra of SNe~Ia (e.g., SN 2011fe, SN 2014J, Bikmaev et al. 2015). Spectra of ASASSN-15nx indeed show 5890\,\AA\ emission (Fig.1, emission "a") that might belong to Co\,III. The problem with 
the Co\,III attribution is that emission does not show the expected significant 
decline with respect to Fe\,II blend in the range of 5000-5500\,\AA\ in the spectrum on day 262 
(Bose et al. 2018) due 
to the radioactive decay of $^{56}$Co, --- the effect always being present in SNe~Ia.
This obvious inconsistency implies that the radioactive mechanism for the ASASSN-15nx should be rejected.

In fact, the 5890\,\AA\ emission  belongs to the Na\,I doublet that scatter the 
continuum and He\,I 5876\,\AA\ radiation; the latter line is scattered by sodium 
doublet due to the reddening in the comoving frame. This effect 
is included in the shown synthetic spectrum (Fig. 1). The adopted ratio of emissivities 
$j(5876)/j(7065)$ is 1, 1, and 2 in spectra on days 53, 87, and 165 respectively.

\subsection{Absence of circumstellar interaction}

Spectra of ASASSN-15nx reveal the Ca\,II 3934, 3968 \AA\ broad absorption. 
This line on day 53 is shown in Fig. 1 (panelinset) together with the model profile. The model is specified 
by the emissivity distribution, the Sobolev optical depth, and the resonant photon loss probability 
$\epsilon_{13} = A_{32}\beta_{23}/(A_{31}\beta_{13} + A_{32}\beta_{23})$, where  $A_{ki}$ 
is the rate of the spontaneous emission, $\beta_{ik}$ is the local escape probability, and  
indices 1, 2, and 3 denote $^2$S, $^2$D, and $^2$P therms. The doublet 3950\,\AA\ corresponds to 
1-3 transition while triplet 8600\,\AA\ corresponds to 2-3 transition.
For optically thick lines the $\epsilon_{13}$ value depends on the excitation 
temperature; for $T = 6000$\,K (cf. Bose et al. 2018) $\epsilon_{13} = 0.7$. This value provides a significant conversion of 3950\,\AA\ doublet emission into the 8600\,\AA\ triplet emission. 
The doublet emissivity therefore can be set comparable to the intensity of the Ca\,II infrared triplet. 
However on day 53 the spectrum in the range of 8600\,\AA\ is absent. We therefore make use of the spectrum on day 87 in which the fluxes of Ca\,\AA\ triplet and H$\alpha$ are comparable. 
Basing on this fact the doublet emissivity on day 53 is also set equal to that of H$\alpha$.
 
The presence of the deep absorption of Ca\,II doublet suggests that the background quasi-continuum 
forms in the inner zone of the envelope. This in turn indicates that the source of the 
ASASSN-15nx luminosity resides in the envelope interiors and not the outer layers. It is the latter that 
is expected in the mechanism of the CS interaction. The internal location of the luminosity 
source thus rules out the luminosity mechanism based on the ejecta collision with the 
dense CS envelope.

%---------------------------------------------------------------

\begin{table}[t]

\vspace{6mm}
\centering
{{\bf Table 1.} Model parametrs for [O\,I] 6300, 6364 \AA\ emission.}
\label{tab:spar} 

\vspace{5mm}\begin{tabular}{l|c|c|c|c|c|c} 
\hline
Model & $v_1$  & $v_2$ & $i$    & $\mu_0$  & $\chi$  & $A$ \\
\hline
ER     & 1500   &  3200  &  70   &  0.6     &    6    & 1  \\
PC1    & 1400   &  2800  &  20   &  0.6     &    2.8  & 0.38  \\
PC2    & 1300   &  2900  &  19   &  0.5     &    2.8  & 0.7  \\
PC3    & 1400   &  2800  &  19   &  0.5     &    2.8  & 0.6  \\
\hline

\end{tabular}
\end{table}
%---------------------------------------------------------------

\subsection{Oxygen and helium abundance}

The high intensity of the oxygen triplet O\,I 7774 \AA\ relative to H$\alpha$ on day 53, 
$F(7774)/F(\mbox{H}\alpha)\approx 0.5$, indicates high oxygen abundance.
Indeed, similar ionization potentials of hydrogen and oxygen secure their comparable ionization, 
so for the comparable intensity of their recombination lines the 
total number of atoms of these elements should be relatively close. Taking into account 
the effective recombination coefficient for these lines (Pequignot et al. 1991) at 
the electron temperature of 6000\,K (Bose et al. 2018) and assuming that oxygen and 
hydrogen are not mixed we find the mass of ionized hydrogen and oxygen to be 
0.03\msun\ and 0.12\msun\ respectively. The ionization fraction cannot be reliably 
estimated. However, the presence of neutral Na suggests that the ionization fraction 
is far from unity. For the ionization fraction of 0.5 the hydrogen and oxygen mass 
are 0.06\msun\ and 0.24\msun\ respectively. Despite these estimates are crude, they 
indicate that the hydrogen mass is low ($\sim 0.1$\msun) and the oxygen mass is relatively high, 
at least several tenths of solar mass.

The intensity of a single He\,I 7065 \AA\ line with respect to H$\alpha$ is 
approximately equal to 0.1. This is twice as high compared to the case of 
normal He abundance and similar ionization fraction of the hydrogen and helium. 
In reality, the helium ionization fraction is likely lower than that of 
hydrogen, so the helium abundance relative to hydrogen is higher.

\subsection{Asymmetry of oxygen distribution}

The profile of the observed [O\,I] 6300, 6364 \AA\ doublet on day 165 significantly 
differs from the synthetic one in the range of radial velocities $|v_r| < 2000$\kms 
(Fig. 1): the observed profile shows flat top with two peaks.
The similar profile with even more pronounced peaks is seen in the spectrum 
on day 262 (cf. Bose et al. 2018). This sort of the oxygen doublet profile is 
inconsistent with the spherical picture and obviously  
related to the asphericity of the oxygen distribution.

To get an idea of the oxygen distribution I consider a model in which on the 
background of the spherically symmetric distribution of emissivity 
$j_s \propto 1/[1 + (v/v_0)^k]$ there is an additional non-spherical component, 
either the equatorial ring 
or bipolar caps with constant contrast $\chi = j/j_s$ in the velocity range $v_1 < v < v_2$. 
Angular size of these components is set by the cosine of the polar 
angle $\mu_0$: $|\mu| < \mu_0$ for the equatorial ring and $|\mu| > \mu_0$ for 
the polar caps. The orientation is set by the inclination angle $i$. The adopted 
doublet ratio is 1:3 that corresponds to optically thin lines. Note, the  
finite optical depth of these lines worsen any axial model.

The modelling shows that the equatorial ring cannot reproduce the profile.
The optimal model of equatorial ring (ER) for the line on day 165 (Fig. 2) is 
characterized by parameters given in Table 1. The ring is assumed to be axisymmetric, which 
corresponds to $A = 1$, where $A$ is the parameter of the azimuthal asymmetry, i.e., the ratio of emissivities in the near and rear hemisphere. In fact, the profile indicates 
the azimuthal asymmetry. However, to reproduce this by the ring model one needs to suppress ring emission not 
only in the far hemisphere, but also at the limb. This sort of a deviation from the axial symmetry is convenient to describe by asymmetric polar caps (PC). 
On day 165 the model 
PC1 (Fig. 2, Table 1) reproduces the profile with the asymmetry parameter 
$A = j(\mbox{red})/j(\mbox{blue}) = 0.38$. On day 262 we show two version: with the constant 
continuum and the inclined continuum (PC2 and PC3 respectively). Models are almost identical which 
emphasises the robustness of the conclusion on the asymmetry of the oxygen distribution.

The fact that the significant fraction of the oxygen shows conspicuous angular and central asymmetry 
at the velocities 1300-3000\kms\ evidences for the significant explosion asymmetry that 
essentially affects oxygen ejecta. This sort of phenomena was never observed in SNe~II, 
although some asymmetry in oxygen doublet related to the $^{56}$Ni asymmetry
was detected in SN~2004dj (Chugai et al. 2005).

\section{Light curve}

After discarding radioactivity and CS interaction, we consider alternative possibilities: 
(i) rotational energy of a young magnetar and (ii) supercritical accretion onto a black hole.
Magnetar mechanism with the luminosity defined by the formula of magneto-dipole radiation 
was employed earlier (Kasen \& Bildsten 2010) for the interpretation of superluminous 
supernovae (SLSN). A problem with the application of this mechanism to ASASSN-15nx is 
evident: the power law decay of the magnetar luminosity is inconsistent with 
the observed exponential luminosity decay. The supercritical accretion onto a black hole 
generally cannot be rulled out. This mechanism was employed for the 
supernova iPTF-14hls (Arcavi et al. 2017, Chugai 2018). The exponential 
luminosity decay is expected, if the accretion rate declines exponentially. It is, however, 
not clear what physics might provide the required evolution of the accretion rate.

Yet another mechanism is conceivable that could account for the exponential luminosity decay.
Suppose that the explosion of ASASSN-15nx was accompanied by the formation of a rapidly rotating 
neutron star with a strong magnetic field but with a relatively low magnetar luminosity.
The process that might provide powerful energy release is the accretion of the gravitationally 
bound gas onto the rotating magnetosphere of the neutron star with mass $M$ and the rotation 
frequency $\omega$. The interaction of the magnetosphere with the accreating gas in the regime of 
a supersonic propeller could power the ASASSN-15nx luminosity.
This scenario requires that for the magnetosphere with the magnetic momentum $\mu$ 
and the accretion rate $\dot{m}$ the magnetosphere radius $r_m = (\mu^4/8\dot{m}^2GM)^{1/7}$ 
should be less than the radius of the light cylinder $r_{lc} = c/\omega$ and be larger than the corotation radius $r_c = (GM/\omega^2)^{1/3}$ (Shakura 1975, Davis et al. 1979). The maximum rate of the rotational energy loss is $L_p = (1/2)\dot{m}(r_m\omega)^2$ (Davis et al. 1979). 
With this luminosity the braking of the neutron star with the inertia moment $I$ is described by 
the equation (Shakura 1975):
\begin{equation}
I\omega\dot{\omega} = -(1/2)\dot{m}(r_m\omega)^2\,.
\label{eq:sdown}
\end{equation}
This equation shows that the constant accretion rate $\dot{m} = const$ implies the exponential 
braking $\omega = \omega_0\exp{(-bt)}$, where $b = 0.5\dot{m}r_m^2/I$.
The rate of the rotational energy loss follows the exponential law 
$L \propto \omega ^2 \propto \exp{(-2bt)}$ which could account for the linear light curve of 
ASASSN-15nx. The steady state accretion rate is thus necessary condition for the 
exponential decay of the luminosity in the proposed mechanism. It is noteworthy that the 
study of the fall-back after the supernova explosion admits a possibility of the 
stationary accretion regime on the time scale of one year (Chevalier 1989).

The complicated transformation of the energy generated by the propeller mechanism is 
left beyond the proposed scenario. It is conceivable that inside the expanding SN envelope 
3-dimensional picture emerges that includes accretion flow and ejected plasma. The latter 
forms hot bubble ($T \sim  5\times10^9$\,K). The energy is transferred in the supernova 
envelope by X/$\gamma$-rays emitted by the hot bubble, although relativistic 
particles can also contribute. In the light curve modelling we assume that all the power 
of the propeller mechanism is deposited in the supernova envelope.
The Table 2 contains the optimal set of parameters of the light curve model: 
the radius of the neutron star, the inertia moment, the magnetic moment, the initial rotation
period, and the accretion rate. The magnetic moment corresponds to the equatorial magnetic 
field at the neutron star surface of $3\times10^{13}$ G. The total mass involved in the accretion during 250 days is $1.6\times10^{-3}$\msun \footnote{The mass of gravitationaly
bound ejecta can attain $\sim 0.1$\msun\ (Chevalier 1989).}.

The present model does not depend on the mass $M_{ej}$ and kinetic energy $E_k$ of the 
supernova envelope. To get an idea about their values we consider a model of 
homogeneous ejecta with the radiation diffusion that is aimed at the description of 
the rising part of the light curve. The optical bolometric luminosity is calculated 
via the thermal energy $E$ and the radiation escape time $t_e$ as $L_{bol} = E/t_e$. The 
time $t_e$ is taken to be equal the average escape time for the photon emitted in 
the center of the homogeneous envelope $t_e = \tau R/2c$ (Sunyaev \& Titarchuk 1980), 
where $R$ is the enevelope radius, $\tau$ is optical depth, and $c$ is the speed of light.
The evolution of the thermal energy is determined by the energy equation 
\begin{equation}
dE/dt = -E/t - E/t_e + L_p\,. 
\end{equation}
The first term is adiabatic losses, the second is the bolometric luminosity, and the third is 
the luminosity of the propeller mechanism. The opacity is assumed to be Thomson opacity 
with the number of free electrons per barion $y_e = 0.2$. This value is supported by the 
following arguments. The envelope with the mass of 1\msun, the kinetic energy 
of $10^{51}$ erg, and mass fraction of H, He, O respectively beng equal 
$x_1 = 0.1$, $x_2 = 0.1$, and $x_8 = 0.8$ at the maximum (14 d) is characterized by the equilibrium 
ionization $y_e = 0.16$ for the effective temperature 10400\,K. For 
somewhat different composition $x_1 = 0.1$, $x_2 = 0.4$, and $x_8 = 0.5$ we get 
$y_e = 0.18$. Allowing for 
deep layers, where the ionization is higher, the value $y_e = 0.2$ seems to be quite 
reasonable. 

Apart from data on the bolometric luminosity (Fig. 3)
we plot also photometry in $V$-band (Bose et al. 2018) that are matched to the bolometric 
light curve at the maximum.
These additional data permit us to get an idea of the rising part of the light curve.
We show two models (Fig. 3) with the different ejecta mass: 3\msun\ with the kinetic energy  
 $E_k = 1.7\times10^{52}$ erg and 0.5\msun\ with the energy $E_k= 1.5\times10^{50}$ erg. 
 Both models well reproduce the light curve. Note, the model light curve model does not include 
 the narrow initial peak related to the shock breakout.

 The uncertainty of the choice of the ejecta mass is eliminated if we take into account the  expansion velocity. The evolution of the photosphere radius at 
 the early stage 
 (Bose et al. 2018) suggests the velocity at the photosphere $\approx 10^4$\kms. 
 This value coinsides with the maximum velocity in the H$\alpha$ wings on day 53.
 The models shown in Fig. 3 are characterised by the maximum velocity of 31000\kms\ in the 
 case of $M_{ej} = 3$\msun\  and 7100\kms\ in the case of $M_{ej} = 0.5$\msun. Both velocities 
 differ from the observational value $\approx 10^4$\kms\ in either side.
 The acceptable mass therefore should lie in the range $0.5 < M_{ej} < 3$\msun. 
 The optimal model with the maximum velocity of 10300\kms is characterized by the ejecta mass 
 $M_{ej} = 0.7$\msun\ and the kinetic energy $E_k = 4.5\times10^{50}$ erg.
 
 Despite adopted approximations, the model of the early light curve indicates 
 that the ejecta mass is low, $\sim 1$\msun\ and the kinetic energy is
  $E_k = (0.5-1)\times10^{51}$ erg. Note that the constraints for the magnetosphere 
  radius $r_c < r_m < r_{lc}$ hold for the considered model.

 %---------------------------------------------------------------

\begin{table}[t]

\vspace{6mm}
\centering
{{\bf Table 2.} Parameters of the light curve model.}
\label{tab:lcpar} 

\vspace{5mm}\begin{tabular}{l|c|c} 
\hline
Parameter   & Units                         &    Value \\
\hline
$R_{ns}$    &    km                           &   12    \\
$I$         &  $10^{45}$ g\,cm$^2$            &   1     \\
$\mu$       &  $10^{30}$\,G cm$^3$           &  51.8   \\
$P_0$       &     с                           &  0.011   \\
$\dot{m}$   & $10^{23}$ g\,s$^{-1}$            &  1.5    \\

\hline

\end{tabular}
\end{table}
%---------------------------------------------------------------

\section{Discussion and conclusions}

The spectra of the unusual supernova ASASSN-15nx with long linear light curve 
permits us to rule out the radioactive mechanism of the luminosity and the mechanism of 
the shock interaction with the dense CS envelope. The alternative mechanism proposed here 
suggests that the neutron star with strong magnetic field and fast rotation lose its rotational 
energy due to the interaction of the magnetosphere with the gravitationally bound ejecta in the 
propeller regime. In the case of the stationary accretion rate, the rotation energy 
loss decreases exponentially thus explaining the linear light curve. Deviations from the 
steady state accretion regime should bring about deviations from the linearity of the light 
curve which could be observed in other supernovae of this category.

The modelling of the initial rising stage of the light curve  
along with the ejecta expansion velocity lead to the ejecta mass estimate which turns out 
low, about 1\msun. Remarkably, the low mass is also indicated by the behavior of $B-V$ color. 
Accoding to data (Bose et al. 2018) $B-V$ rises the same way as in the case of SNe~IIP. 
However, if for SNe~IIP the rise span is ~$\approx 100$ days, in the case of ASASSN-15nx 
the rise takes only $\approx 50$ days. Since the $B-V$ behavior reflects the envelope cooling 
in the photospheric regime, this means that for ASASSN-15nx the duration of the 
photospheric phase is twice as short compared to SN~IIP. Given comparable 
expansion velocities this indicates significantly lower mass of ASASSN-15nx compared to 
SNe~IIP. Combining the neutron star and ejecta we come to the preSN mass 
in the range of $\sim 2-2.5$\msun. Noteworthy that the hydrogen mass in the preSN was likely 
less then 10\%.

A difficult question arises concerning the origin of ASASSN-15nx. In a single progenitor scenario 
the preSN should be a helium core with the leftovers of the hydrogen envelope. For the 
He core mass of 2.5\msun\ the preSN presumably is the outcome of the evolution of 
a star with ZAMS mass about 10\msun\ (Nomoto 1984). This scenario, however, does not 
admit the significant amount of the oxygen over the collapsing core, which contradicts to the 
presence of at least several tenth of solar mass of synthesised oxygen in ASASSN-15nx.

An alterantive evolutionary scenario for ASASSN-15nx suggests close binary that 
consists of ONeMg white dwarf (primary) and low mass star with CO degenerate core at the 
AGB stage. The evolution of this system might result in the merging of CO-core and ONeMg white dwarf with the subsequent collapse of the ONeMg white dwarf initiated by $e$-capture.
The advantage of this scenario is that it could account for the presence of a significant 
amount of oxygen in low mass ejecta. The oxygen in this scenario is the result of the 
disruption of the CO-core of the secondary component during the merger. 
The remains of the hydrogen envelope 
of the secondary could explain the presence of the low amount ($\sim0.1$\msun) of hydrogen 
in ASASSN-15nx ejecta. The merging scenario and related fast rotation could be the reason for the explosion asphericity indicated by the oxygen doublet.

\section{Acknowledgement}

I thank Konstantin Postnov, Lev Yungelson, Victor Utrobin for discussions, and  
Subo Dong for sharing the spectra of ASASSN-15nx.

\pagebreak   
%****************************************************************

\pagebreak   
%****************************************************************

%xxxxxxxxxxxxxxxxxxxxxxxxxxxxxxxxxxxxxxxxxxxxxxxxxxxxxxxxxxxxxx
\clearpage
%xxxxxxxxxxxxxxxxxxxxxxxxxxxxxxxxxxxxxxxxxxxxxxxxxxxxxxxxxxxxxx
\begin{figure}[h]
	\epsfxsize=19cm
	\hspace{-2cm}\epsffile{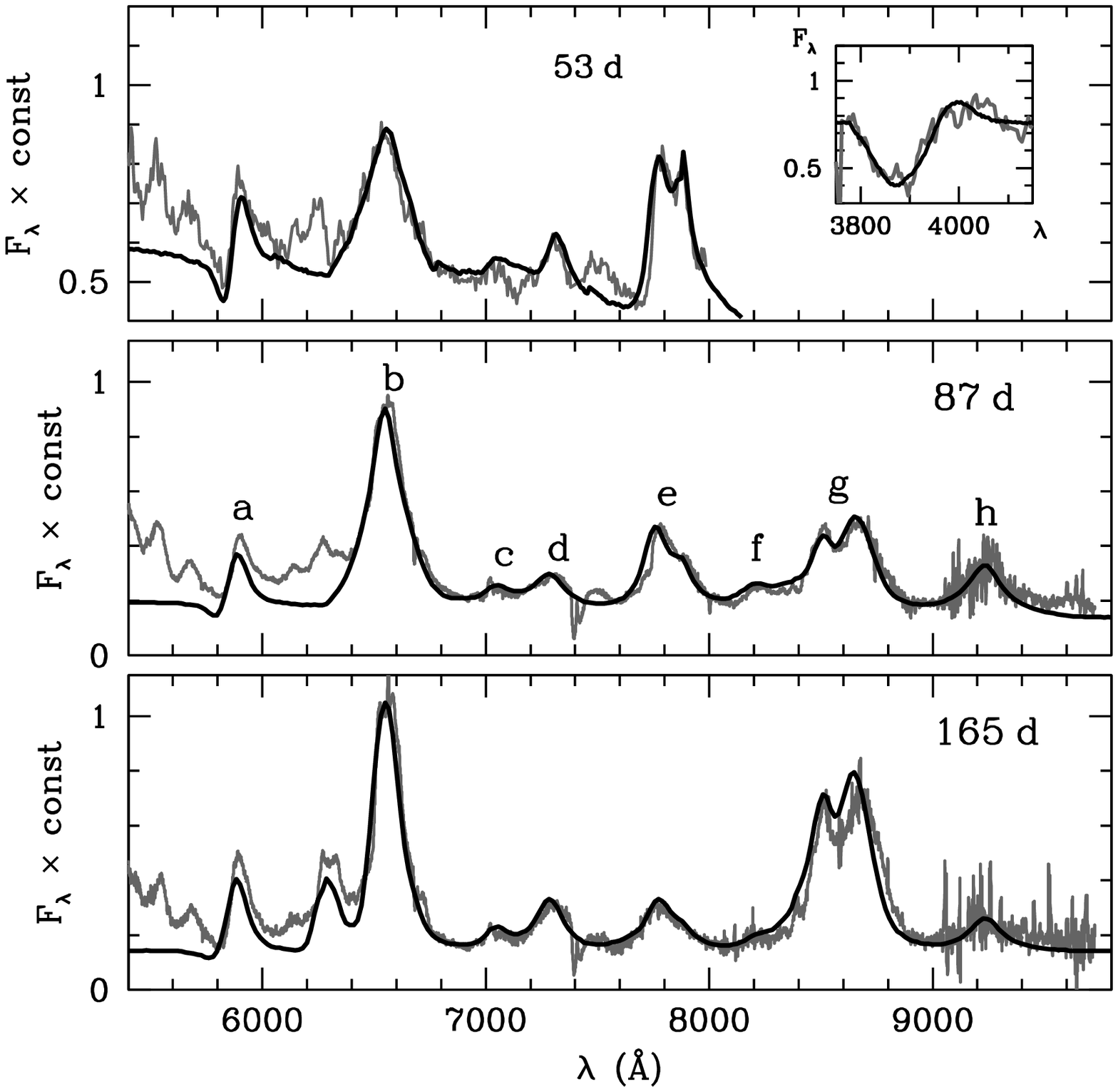}
	\caption{\rm  
	Observed spectra of ASASSN-15nx ({\it grey}) for three epochs (53, 87, 165 days) compared to 
	synthetic spectra. On day 53 the inset shows doublet Ca\,II H,K. For convenience on day 87 
	principal emission features are marked by letters in alphabetical order. The profile of 
	[O\,I] 6300, 6364 \AA\ on day 165 with two peaks on top reflects aspherical oxygen distribution 
	(see text).
	}
\end{figure}
%========================================================

\clearpage
%xxxxxxxxxxxxxxxxxxxxxxxxxxxxxxxxxxxxxxxxxxxxxxxxxxxxxxxxxxxxxx
\begin{figure}[h]
	\epsfxsize=19cm
	\hspace{-2cm}\epsffile{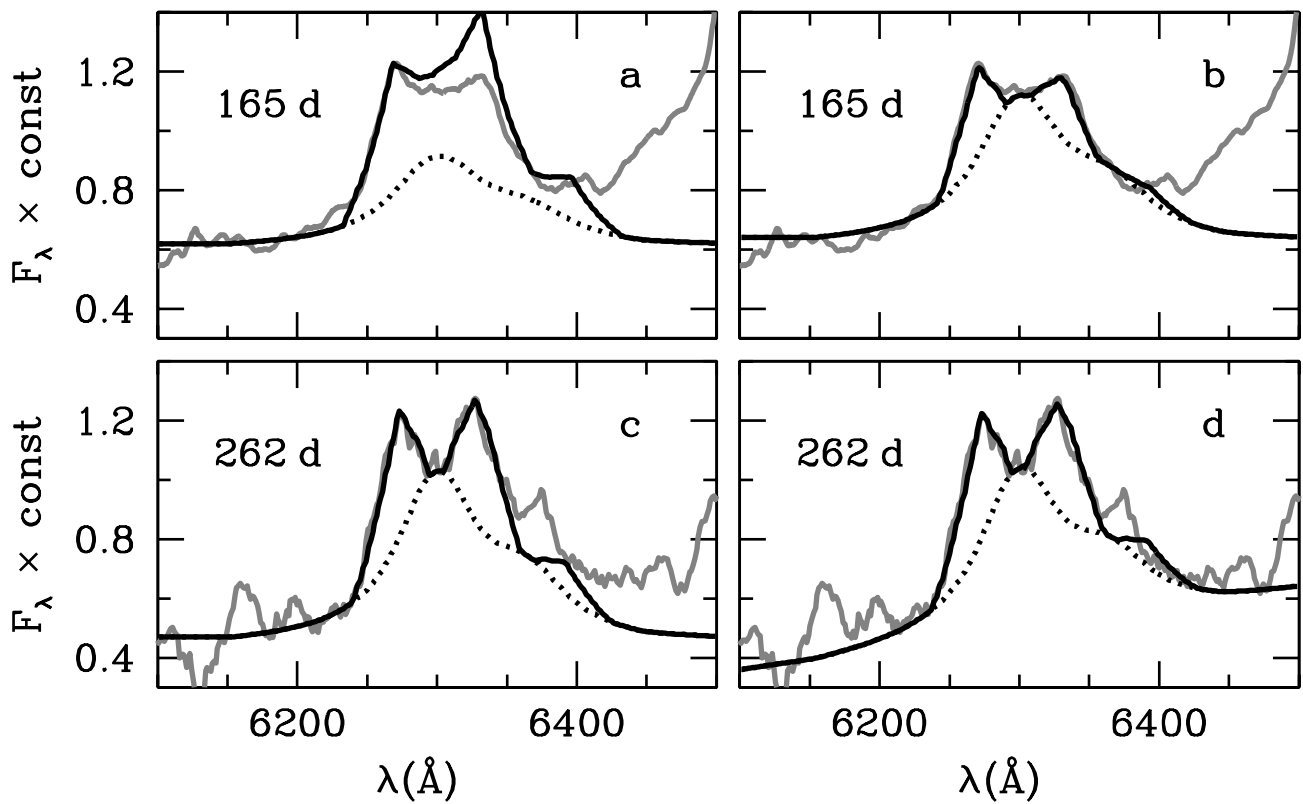}
	\caption{\rm  
	Observed oxygen doublet [O\,I] 6300, 6364 \AA\ on days 165 and 262 ({\it grey}) with 
	overplotted model profiles. {\it Dotted} line shows the contribution of the 
	spherical component. The panel {\bf a} shows the model of equatorial ring (ER, Table 1);
	this model does not fit the observed profile on day 165. The same observed profile on 
	the panel {\bf b} is well reproduced in the model of asymmetric polar caps (PC1, Table 1).
	The panel {\bf c} shows the model PC2 (Table 1) that fits the observed profile on day 262, 
	and the panel {\bf d} shows the same observed profile and the model PC3 (Table 1) with the
	inclined continuum.
		}
\end{figure}
%========================================================

\clearpage
%xxxxxxxxxxxxxxxxxxxxxxxxxxxxxxxxxxxxxxxxxxxxxxxxxxxxxxxxxxxxxx
\begin{figure}[h]
	\epsfxsize=19cm
	\hspace{-2cm}\epsffile{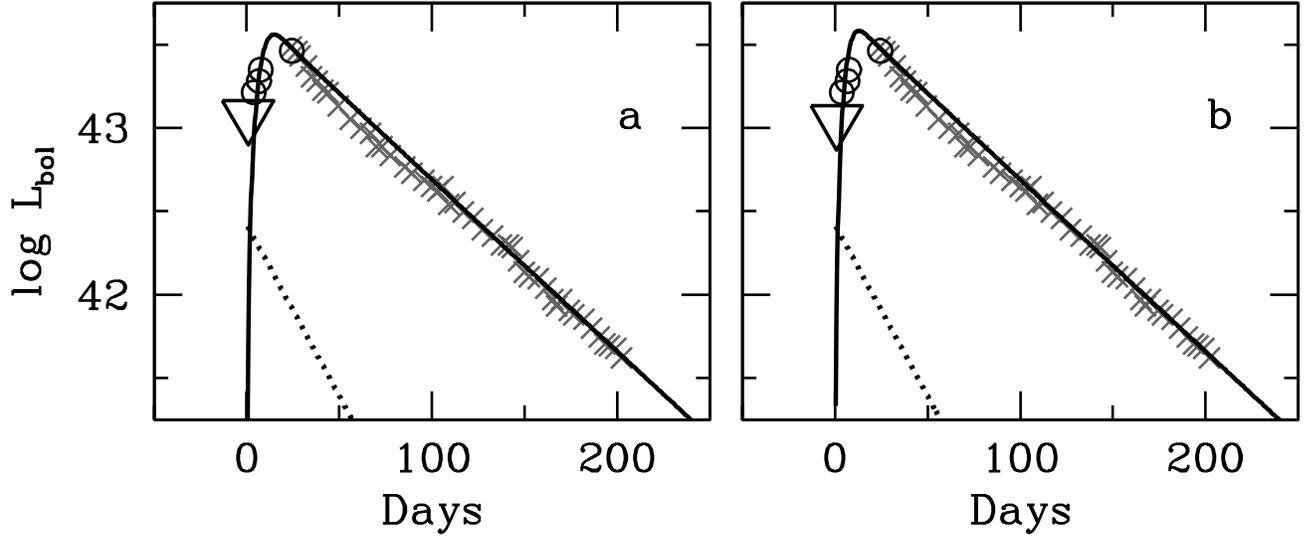}
	\caption{\rm  
	Bolometric light curve of ASASSN-15nx ({\it crosses}) and the model ({\it solid} line) 
	for two versions: ejecta mass 3\msun (panel {\bf a}) and 0.5\msun (panel {\bf b}). 
	{\it Circles} are the photometric data in $V$ band normalised on the early bolometric 
	luminosity. {\it Triangle} symbol corresponds to the upper limit in $V$ band. 
	Magnetar luminosity is shown by the {\it dotted} line.
	}
\end{figure}
%========================================================

\end{document}